\documentclass[twocolumn,showpacs]{revtex4}

\usepackage{graphicx}% Include figure files
\usepackage{dcolumn}% Align table columns on decimal point
\usepackage{bm}% bold math

\input epsf

\begin{document}

\title{\Large Matter in the Bulk and its Consequences on the Brane:
A Possible Source of Dark Energy}

\author{\bf~Subenoy~Chakraborty$^1$\footnote{schakraborty@math.jdvu.ac.in},
~Asit~Banerjee$^2$~and~Tanwi~Bandyopadhyay$^1$}

\affiliation{$^1$Department of Mathematics,~Jadavpur
University,~Kolkata-32, India.\\ $^2$Department of
Physics,~Jadavpur University,~Kolkata-32, India.}

\begin{abstract}
The usual brane world scenario with anti de Sitter bulk has been
generalized by considering a general form of energy momentum tensor
in the bulk. The modified Einstein equation on the brane has been
constructed. Two examples have been cited of which, the first one
shows the usual brane equations when matter in the bulk is a
negative cosmological constant. In the second example, the bulk
matter is in the form of perfect fluid and as a result, an effective
perfect fluid is obtained in the brane. Also it is noted that the
effect of the dust bulk on the brane shows a dark energy behaviour
and may be a possible explanation of the dark energy from the
present day observational point of view.
\end{abstract}


\maketitle

Branes play a fundamental role in the string theory, especially
the D-branes on which open strings can end. The open strings
describing the non gravitational sector are supposed to be
attached to the branes at their end points, while the closed
strings on the gravitational sector can propagate in the bulk,
which is the higher dimensional embedding spacetime [1]. However,
drastic simplifications are usually made regarding the
localization of matter field on the brane with the aim to
establish consistency of the observational predictions. According
to the standard brane world idea, the particles are confined on a
hypersurface known to be the brane, embedded in a higher
dimensional anti de-Sitter spacetime called the bulk. The scalar
field [2] has further been included in the bulk in order to
generalize the so called Randall-Sundrum model [3].\\

On the other hand, from the phenomenological point of view, the
matter in the bulk can be anything and is supported by the existence
of a higher dimensional embedding space. The existence of bulk
matter may modify the cosmological evolution on the brane to a great
extent and may lead to possible explanation of the present day
observations. It may be speculated that though the matter in the
bulk [4] has no exotic features, yet its gravitational attributes in
the brane is a major source of dark energy. Then, as a result of
this bulk matter and $Z_{2}$ symmetry, the divergence of the brane
energy-momentum tensor is no longer zero [5,~6]. This non
conservation of brane energy-momentum tensor reflects an exchange of
energy momentum between the bulk and the brane. In case of vacuum or
(negative) cosmological constant in the bulk, there is no such
exchange and hence the Bianchi identity simply imposes a constraint
on the projected Weyl tensor on the brane [7,~8]. In this letter, we
attempt to outline the procedure for arriving at the final form of
the field equations in the brane along with the effect induced by
the bulk containing the most general form of matter. This effort
will hopefully widen the scope of studying the brane world
models in a more general context.\\

Suppose the five dimensional metric on the bulk is written as

\begin{equation}
ds^{2}=g_{AB}~dy^{A}dy^{B}
\end{equation}

and the brane is defined to be the hypersurface $y^{5}$=constant,
(say zero). So $n_{A}=\delta^{5}_{A}$ is the unit normal space
like vector on the brane hypersurface. Assuming $g_{55}=1$ [5],
the induced metric on the brane can be written as

\begin{equation}
q_{AB}=g_{AB}-n_{A}n_{B}
\end{equation}

Thus a natural choice of coordinates is $y^{A}=(x^{\mu},y^{5})$
with $x^{\mu}=(t,x^{i})$ as the space-time coordinates on the
brane. The extrinsic curvature orthogonal to $n^{A}$ is given by\\

~~~~~~~~~~~~~~~~~~~~$K_{AB}=\frac{1}{2}~
\mathcal{L}_{\overrightarrow{n}}~g_{AB}$\\

Now, using the five dimensional Einstein equation on the bulk

\begin{equation}
G_{AB}=\kappa_{5}^{2}\left[T_{AB}+S_{AB}~\delta(y^{5})\right]
\end{equation}

the Einstein equation on the brane can be written as [5,~6]

\newpage

\begin{eqnarray*}
G_{\mu\nu}=\frac{2}{3}\kappa_{5}^{2}\left[T_{AB}q_{\mu}^{A}q_{\nu}^{B}
+\left(T_{AB}n^{A}n^{B}-\frac{1}{4}T\right)q_{\mu\nu}\right]
\end{eqnarray*}
\begin{equation}
-E_{\mu\nu}+KK_{\mu\nu}-K_{\mu}^{\sigma}K_{\nu\sigma}-\frac{1}{2}q_{\mu\nu}(K^{2}
-K^{\alpha\beta}K_{\alpha\beta})
\end{equation}

where $T_{AB}$ is the energy-momentum tensor on the bulk,
$S_{\mu\nu}$ is the effective energy-momentum tensor localized on
the brane and $\delta(y^{5})$ stands for the localization of brane
contribution.\\

Also, $E_{\mu\nu}=C^{A}_{BDF}n_{A}n^{D}q_{\mu}^{B}q_{\nu}^{F}$,
is the projection of the bulk Weyl tensor orthogonal to $n^{A}$,
with $E_{[\alpha,\beta]}=0 =E_{\alpha}^{\alpha}$. The equation
(4) in fact follows from the Gauss equation and the decomposition
of the Riemann tensor into
Weyl curvature, the Ricci tensor and the Ricci scalar.\\

Also the Codazzi equation has the form

\begin{equation}
\nabla_{\nu}K_{\mu}^{\nu}-\nabla_{\mu}K=\kappa_{5}^{2}~T_{AB}~n^{A}~q_{\mu}^{B}
\end{equation}

Further, remembering Israel's junction condition [5,6,8,9] on the
brane hypersurface (assuming the brane to be infinitely thin), we
get

\begin{equation}
[q_{\mu\nu}]=0~~~\text{and}~~~[K_{\mu\nu}]=-\kappa_{5}^{2}~(S_{\mu\nu}
-\frac{1}{3}~S~q_{\mu\nu})
\end{equation}

where $[f]=\displaystyle{\lim_{y^{5}\rightarrow0^{+}}}~f-
\displaystyle{\lim_{y^{5}\rightarrow0^{-}}}~f=f^{+}-f^{-}$.\\

Now imposing $Z_{2}$ symmetry on the bulk spacetime with brane as
the fixed point we have

\begin{equation}
K_{\mu\nu}^{+}=-K_{\mu\nu}^{-}=-\frac{1}{2}~\kappa_{5}^{2}~(S_{\mu\nu}
-\frac{1}{3}~S~q_{\mu\nu})
\end{equation}

(henceforth neglecting the indices $\pm$ for simplicity)\\

Then from equation (4) using (7), the form of the effective
Einstein equation on the brane takes the form

\begin{equation}
G_{\mu\nu}=\kappa_{4}^{2}~\tau_{\mu\nu}+\kappa_{5}^{4}~\Pi_{\mu\nu}-E_{\mu\nu}
+T_{\mu\nu}^{(P)}
\end{equation}

where

\begin{eqnarray*}
T_{\mu\nu}^{(P)}=\frac{2}{3}~\kappa_{5}^{2}\left[T_{AB}~q_{\mu}^{A}~q_{\nu}^{B}
+\left(T_{AB}n^{A}n^{B}-\frac{1}{4}T\right)q_{\mu\nu}\right.
\end{eqnarray*}
\begin{equation}
\left.-\frac{1}{8}~\kappa_{5}^{2} ~\lambda^{2}~q_{\mu\nu}\right]
\end{equation}

is termed as the projected part of the bulk energy-momentum
tensor on the brane.\\

Here we have decomposed $S_{\mu\nu}$ as

\begin{equation}
S_{\mu\nu}=-\lambda~q_{\mu\nu}+\tau_{\mu\nu}
\end{equation}

with $\lambda~(>0)$ and $\tau_{\mu\nu}$, the vacuum energy (brane
tension) and the energy-momentum tensor respectively in the brane.
The $\Pi_{\mu\nu}$ term, which is quadratic in $\tau_{\mu\nu}$, has
the usual expression [5,~8]

\begin{equation}
\Pi_{\mu\nu}=\frac{1}{12}~\tau~\tau_{\mu\nu}-\frac{1}{4}~\tau_{\mu\alpha}~\tau_{\nu}^{\alpha}
+\frac{1}{24}~q_{\mu\nu}~[3~\tau_{\alpha\beta}~\tau^{\alpha\beta}-\tau^{2}]
\end{equation}

and ~~~~$\kappa_{4}^{2}=\frac{1}{6}~\kappa_{5}^{4}~\lambda$.\\

As a consequence of the Codazzi equation and the $Z_{2}$
symmetry, the divergence of the brane energy-momentum tensor is
related to the projection of the bulk energy-momentum tensor by
the relation [6]

\begin{equation}
\nabla_{\nu}~\tau_{\mu}^{\nu}=-2~T_{AB}~n^{A}~q_{\mu}^{B}
\end{equation}

Thus equation (8) is the effective Einstein equation on the brane
with matters in the bulk and brane, characterized by the
energy-momentum tensors $T_{AB}$ and $\tau_{\mu\nu}$ respectively.
Note that, these matters are not completely arbitrary but are
restricted by the relation (12) [Also the Bianchi identity
$\nabla_{\nu}G_{\mu}^{\nu}=0$ gives one more restriction relating
the divergence of $E_{\mu\nu},~\Pi_{\mu\nu}$ and $T_{AB}$ and is
complicated in nature ].\\

We shall now cite below a few examples as special cases:\\

I. \underline{Bulk as the anti de-Sitter space time}~:\\

Here the energy-momentum tensor in the bulk is

\begin{equation}
T_{AB}=-\Lambda~g_{AB}
\end{equation}

$\Lambda$ being the five dimensional cosmological constant.\\

Now using (13), the effective Einstein equation (8) simplifies to

\begin{equation}
G_{\mu\nu}=\kappa_{4}^{2}~\tau_{\mu\nu}+\kappa_{5}^{4}~\Pi_{\mu\nu}-E_{\mu\nu}
-\Lambda_{(4)}~q_{\mu\nu}
\end{equation}

where~~~
$\Lambda_{(4)}=\frac{1}{2}~\kappa_{5}^{2}\left(\Lambda+\kappa_{5}^{2}
~\frac{\lambda^{2}}{6}\right)$ is the effective cosmological
constant on the brane [5,~6,~8].\\

This is the usual Randall-Sundrum brane model.\\

II. \underline{Bulk matter is in the form of perfect fluid}~:\\

Here the form of the energy-momentum tensor is

\begin{equation}
T_{AB}=(\rho+p)~u_{A}u_{B}+p~g_{AB}
\end{equation}

where the velocity five vector is given by $u_{A}$ and $\rho,~p$
stand for the density and pressure of the perfect fluid in the
bulk. Here $u_{A}$ is the time like vector orthogonal to the
space like normal vector $n_{A}$ i.e, $u_{A}~n^{A}=0$.\\

Thus the projected part of the bulk matter (given by equation
(15)) given in equation (9) simplifies to

\begin{equation}
T_{\mu\nu}^{(P)}=\frac{2}{3}~\kappa_{5}^{2}~[~(\rho_{eff}+p_{eff})~u_{\mu}u_{\nu}
+p_{eff}~q_{\mu\nu}]
\end{equation}

where the effective density $\rho_{eff}$ and the effective
pressure $p_{eff}$ on the brane due to the projection of the bulk
energy-momentum tensor may be written as

\begin{equation}
\rho_{eff}=\frac{3\rho}{4}+\frac{\kappa_{5}^{2}~\lambda^{2}}{8},~~~~~
p_{eff}=p+\frac{\rho}{4}-\frac{\kappa_{5}^{2}~\lambda^{2}}{8}
\end{equation}

We note from the above expressions (17) that the effective density
and pressure on the brane contributed by the bulk matter are
combinations of the bulk energy density, bulk pressure and the brane
tension. An interesting conclusion arising from the above
consideration is that, a pure dust in the 5-D bulk contributes to
both density and pressure of the brane in the form of an effective
perfect fluid. All these contributions from the bulk add to the
already existing brane energy density and pressure arising out of
the brane matter itself. Now\\

~~~~~~~~~~~$\rho_{eff}+p_{eff}=\rho+p$\\

and ~~~~~~$\rho_{eff}+3p_{eff}=\frac{3\rho}{2}+3p
-\frac{\kappa_{5}^{2}~\lambda^{2}}{4}$.\\

Note that for dust filled bulk, if
$\rho<\frac{\kappa_{5}^{2}~\lambda^{2}}{6}$, then strong energy
condition is violated for the induced matter in the brane. But for
perfect fluid in the bulk, if the dominant energy condition is
satisfied, then it will also be obeyed by the induced matter in the
brane. In other words, bulk with dust matter shows its impact on the
brane models as a possible source of dark energy. Therefore,
although we can not probe the bulk, but the cosmological evolution
on the brane is modified by the dark energy through the bulk
gravitational influence. So our scenario of brane dynamics
drastically changes if we include the above additional
contributions, which might be an interesting problem for future work.\\

\end{document}